# Casual Social Media Use among the Youth: Effects on Online and Offline Political Participation

Mehdi Barati

ORCID Nr: 0000-0002-3447-4197
*Department of Information Science, CEHC, State University of New York at Albany, New York, U.S.*
*mbarati@albany.edu*

*Abstract: Previous studies suggest that social media use among the youth is correlated with online and offline political participation. There is also a mixed and inconclusive debate on whether online political participation in the youth increases offline political participation. This study uses three models of OLS, two-way fixed effects, and an instrumental variable approach to make causal inferences about the social media use and online and offline political participation of the youth. The analyses provide evidence of a significant effect of casual social media use on online political participation and no effect or negligible effect on offline political activity and voting behavior. The results from fixed effects and instrumental variable models provide strong evidence of elasticity between online political participation and offline political activity in young individuals. On average, a one percent increase in online political participation increases the offline political activity index by 0.12 percent.*

*Acknowledgment: I am very grateful to Dr. Bahareh Ansari for her insightful comments.*

## 1. Introduction

Social media use has been increasing exponentially during the last decade. According to the Pew Research Center, the social media use rate among American adults increased from only 5% in 2005 to 72% in 2018 (Pew Research Center, 2021). Another survey from the same center found that, in 2018, 95 % of teens had access to a smartphone, and 45% said they are online 'almost constantly' (Anderson & Jiang, 2018). A more recent survey found that the majority of Americans say they use YouTube and Facebook, while use of Instagram, Snapchat and TikTok is especially common among adults under 30 (Auxier & Anderson, 2021). This extensive presence of social media in everyday life of the young population might affect their social and political behaviour. Youth political participation has always been challenging, and the recent social media presence has made it more complicated.

The effect of social media on political participation is twofold. First, new forms of political participation have emerged on online platforms. For example, there are four main forms of online political participation within the current digital media space: people use new media to circulate (e.g., blogging, forwarding, and sharing), collaborate (e.g., Wikipedia), create (e.g., producing media in platforms like YouTube), and connect (e.g., interaction through social media) (Jenkins et al., 2018), all of which could be encouraged by social media use. Second, social media use may be a good channel to encourage real-world political activities (such as voting, attending campaign events, and donating to the candidates) because it provides a platform where political messages are circulated through peers.

Because of their higher use of social media and their higher openness to change and to adapt to the environment, the political participation of the youth is more likely to be influenced by social media use. Specific attention to youth participation in civic and political life is important due to a couple of reasons. First, the young population is more engaged with social media platforms, and therefore they are more dependent on social media for gaining information about any topic, including the political ecosystem. Second, a common perception among the youth is that they have been ignored in conventional forms of political participation (Jenkins et al., 2018). Therefore, they might find few alternatives for accessing the institutionalized forms of political participation, such as political parties or lobbying, that match their interests. The role that social media use could have in reducing political inequality has also been addressed in the literature (Xenos et al., 2014).

In a recent study, Lorenz-Spreen et al. (2022) conducted a systematic literature review of the causal and correlational evidence on the link between digital media use and different political variables, to examine the causal relationship between digital media use and the decline in democracy. They find that most of the extant studies have provided correlational evidence for the association between social media use and political participation that can be beneficial for democracy development in autocracies and emerging democracies. They also documented evidence of detrimental effects on political trust, polarization, and populism in established democracies. Unlike the current study, Lorenz-Spreen et al. (2022) adopted a broad conceptualization of digital media which ranges from any access to internet access to the use of specific social media platforms.

The effects of social media use on individual and group-level behavior have been investigated in different areas. Early researchers had optimistically focused on the positive impacts, given the dramatic decrease in communication and information-sharing costs induced by the spread of social media (Vitak et al., 2011). Recent studies are more concerned with the negative impacts of social media use, especially on individuals' emotional states. Kross et al. (2013) show the adverse effects of Facebook use on subjective well-being. Allcott et al. (2020), by conducting a randomized experiment, examine the impacts of deactivating the Facebook account for the four weeks before the 2018 US midterm election on individuals' voting behavior. They find that deactivating Facebook accounts reduces online activity while increasing offline activities reduces factual news knowledge and political polarization, more importantly, increases subjective well-being.

 

The existing literature on the effect of social media use on young adults' political participation needs to be more conclusive. Boulianne (2015), for example, conducts a meta-analysis of the studies about how social media might affect citizens' participation in civic and political life. She focuses on the studies of the effects of social networking sites and includes quantitative survey-based studies focused on behavioral dependent variables like voting or campaigning. The results show that from 36 studies (170 coefficients), 82% of them were positive. However, only 49% of the coefficients were statistically significant. More importantly, only 11% of the studies use random sampling. Moreover, Boulianne (2015) finds that most studies do not need a clear criterion for defining social media use, which leads to varying significance and magnitude of coefficients among previous studies. Also, many of the previous studies only investigate one specific platform, such as Facebook or Twitter. However, these platforms' design, technology, and business model are constantly changing. Therefore, the estimates of the effect of a specific platform might hold only for a short time.

This study attempts to mitigate the statistical limitations of previous studies by using two waves of a survey among a randomly selected nationally representative sample of the youth. Moreover, this study investigates general social media use instead of a specific platform to provide a generalizable estimate. In terms of definition, this study adopts the well-accepted definition offered by Carr & Hayes (2015). Based on this definition, social media are "Internet-based, distrained, and persistent channels of mass personal communication facilitating perceptions of interactions among users, deriving value primarily from user-generated content. Social media are Internet-based channels that allow users to interact opportunistically and selectively self-present, either in real-time or asynchronously, with both broad and narrow audiences who derive value from user-generated content and the perception of interaction with others."

For conceptual clarity, we adopt the social media definition by Obar & Wildman (2015) with four defining characteristics. Social media platforms are internet-based applications with user-generated content as their main currency. People and individuals can create their profiles on the platforms, and service providers maintain those profiles. Finally, social media platforms facilitate the development of social networks online by connecting existing profiles. The theoretical framework used in this study is the social media political participation model (SMPPM) proposed by Knoll et al. (2020) based on an integration of goal system theory, appraisal theory, and the priming paradigm. SMPPM distinguishes between two types of exposure to political content, namely intentional and incidental. Entertainment motivation and social interaction motivation are among the factors that encourage incidental exposure. Received information, if relevant to participatory goals, make the users implicitly formulate and activate a goal, leading to low-effort online political participation if the goal is perceived as consistent with other dominant goals (Knoll et al., 2020).

Building on SMPPM (Knoll et al., 2020), this study hypothesizes that casual use of social media platforms leads to incidental exposure to political content and the exposure consequently leads to online political participation (low effort participation in the SMPPM framework) through implicit goal formation and activation, and exposure to other sources of information. Increased online political participation, in turn, leads to increased offline political participation (high-effort participation) by increasing the chance of intentional exposure to political content and developing the relevant social network. Figure 1 illustrates our adaptation of the SMPPM model, which builds the conceptual model of this study. Using this conceptual framework, this study tests four hypotheses:

H1: Casual social media use (use of social media to connect with friends and family) affects the extent to which young adults engage in online political participation.

H2: Casual social media use (use of social media to connect with friends and family) affects the extent to which young adults engage in offline political participation.

H3: Casual social media use (use of social media to connect with friends and family) affects the probability of voting among young adults.

H4: Online political participation affects the level of engagement in offline political activities among young adults.

*Figure 1: Conceptual model of the study. Direct and indirect effects of casual social media use*

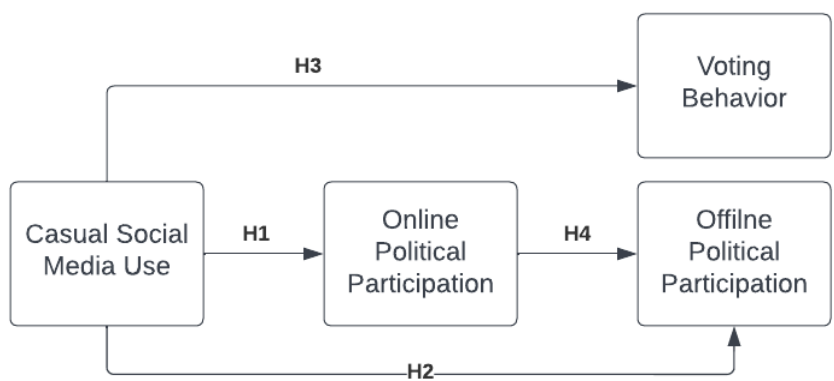

## 2. Data and Methods

### 2.1. Data

This study uses the second and third waves of the Youth Participatory Politics Survey Project (YPPSP), a longitudinal survey of a nationally representative cohort of young adults in the United States conducted by Cathy Cohen at the University of Chicago (Cohen & Kahne, 2018). YPPSP started in 2011 to explore young adults' political attitudes and behaviors. While the first wave of this survey had a cross-sectional design, the second and third waves of this survey repeated the same questions on a cohort of young adults in two time periods. The second wave of the survey was conducted in 2011 with a sample of 1033 young adults ranging from 13 to 25 years old. In the third follow-up wave, all of the 1033 participants were surveyed again, with ages ranging from 17 to 29 years old (Cohen & Kahne, 2018). The questions in the survey were mainly about new media engagement, political and civic attitudes, involvement in communities, political participation, and news sources.

## 2.2. Variables

Independent variable. Casual social media use or the use of social media for communication with friends and family is constructed based on the responses to a set of questions about different methods that respondents use to interact with their friends and family via social media (questions 11-15 in Table 1). Therefore, the independent variable is constructed as the sum of the scores of the frequency of using different social media platforms for five activities that are centered around family and friends. The frequency of the use of each method is measured on a five-point scale ranging from "Never" to "Daily" (i.e., 1= never, 2= less than once a month, 3= at least once a month, 4= at least once a week, 5= daily). The same questions were used in previous research to measure friendship-driven online activity (Kahne & Bowyer, 2018), which is a similar concept to casual social media use in the current study. These questions represent distinct methods that together comprise the range of activities available for communication with friends and family on social media. The independent variable is constructed as the sum of the scores of the frequency of using different social media platforms for five activities that are centered around family and friends. Thus, higher scores of this variable should be interpreted as higher levels of social media use. If a respondent fills 1-1-1-1-1 out all the questions related to casual social media use, then their score for casual social media use will be 0. However, this respondent is not excluded from the regression analyses. But the value of the independent variable for them will be zero.

*Table 1: Survey questions used to measure variables of this study.*

| Casual social media use |
|---|
| People interact with family and friends through a variety of online methods. How often would you say you interact with your friends and family by: |
| Q11. Sending messages, sharing status updates, or chatting online using social network services like Twitter or Facebook (1-5) |
| Q12. Sharing links or forwarding information or media through social network services like Twitter or Facebook (1-5) |
| Q13. Tagging friends and family members in posts, photos, or videos on social networking sites (1-5) |
| Q14. Commenting on something a friend or family member posted on a social networking site (1-5) |
| Q15. Visiting websites and other online material that friends or family members have posted or circulated." (1-5) |
| Online political participation |
| People use a variety of methods to gather and share information about political candidates, campaigns, or political issues. Please tell us how often you have done the following during the past 12 months. |
| Q84. Forwarded, re-tweeted or posted someone else's article, blog, picture or video about a political campaign, candidate, or issue |
| Q85. Created and circulated your own article, blog, picture, or video about a political campaign, candidate, or issue to an online site |
| Q86. Commented online or tweeted about an article, blog, picture, or video you saw about a political campaign, candidate, or issue |
| Q87. Posted a status update or sent an email, Tweet or instant message about a political campaign, candidate, or issue |
| Q88. Followed someone on Twitter for political information, news, or opinions |
| Q89. Signed up to receive information from candidates or campaigns via email or text |
| Offline political participation |
| We find that many people participate in politics in other ways besides voting. Please tell us if you have supported a candidate, political party, or political issue during the past 12 months by: |
| Q77. Attending a meeting, rally, speech or dinner |
| Q78. Working on an election campaign |
| Q79. Wearing a campaign button, putting a campaign sticker on your car, or placing a sign in your window or in front of your home |
| Q81. Donating money to a candidate, party, or political organization |
| Q82. Raising money from your friends and network for a candidate, party, or political organization. |

Dependent variables. Three dependent variables are used in this study: online political participation, offline political participation, and voting behavior. To measure online political participation, this study uses six survey questions that asked respondents about the frequency of their social media use and other online platforms (such as websites and search engines) to interact with and influence a political campaign, candidate, or issue, in the past 12 months (questions 84-89 in Table 1). The frequency of usage is measured on a five-point scale ranging from "Never" to "Several Times a Week" (1= never, 2= less than once a month, 3= at least once a month, 4= at least once a week, 5= daily). Online political participation is measured as the sum of the responses to the six questions. The scale of responses is changed from 1-5 to 0-4, and therefore, the result is an index between 0 to 24. The log scale was decided for online political participation because the study attempted to examine what percent change in political participation should be expected with one unit change in social media use. Change in political participation is more meaningful in percentage terms rather than absolute terms and therefore the use of a log scale makes the interpretation of regression results easier (Bailey, 2016, p. 218).

To measure offline political activity, this study uses the responses to five questions asking about whether or not they participated in political events in the previous 12 months (questions 77-82 in Table 1). Offline Political Activity is measured as the sum of the responses to the above five questions which led to an index between 0 to 5. The index is changed into a log scale to avoid the violation of the linearity assumption. The third dependent variable is voting behavior which is constructed based on a question that asks whether or not the respondents voted in the election last November or expect to vote regularly (if they are under 18).

## 2.3. Empirical challenge

Confounding bias and reverse causality are the two important challenges that this study attempts to resolve. This study uses different techniques to reduce the biases as much as possible and rigorously tests the hypotheses. The frequency of social media use is not randomly assigned to young people. They choose to spend time on social media and their choice might be related to their interest, personality, family background, education level, and other factors. These factors also determine individuals' online and offline political participation. For example, those who live in a family with high levels of disciplinary regulations are less likely to spend several hours a day on social media and more likely to engage in online and offline political participation, compared to those who live in less disciplined families. This makes the estimated coefficient by Ordinary Least Squares (OLS) regression negatively biased (i.e., the estimated coefficient would be less than the true effect size).

On the other hand, those who live in families with a higher level of household income might have more leisure time to spend on social media and also have more opportunities to engage in political activities relative to those who live in lower-income families. In this case, the estimated coefficient would be biased positively. Overall, it is highly probable that the simple OLS model is subject to confounding bias with an unclear direction.

To solve the issue of confounding bias for testing H1 and H2, the current study uses panel fixed-effects and random-effects models. Panel fixed-effects models exploit the change within a unit over time and control for all time-invariant differences between the individual units that could drive the association. Most of the important confounding factors in the context of social media use and political participation are time-invariant factors, such as family background, parents' educations, personality traits, genetic factors, demographics. Moreover, time-variant factors that could confound the OLS estimates are adjusted by measured variables of education level, age, marital status, household income level, household size, and employment status at two time periods. The use of TWFE requires that there are linear additive effects (Imai & Kim, 2021).This is satisfied in this study. Furthermore, TWFE requires that it is equivalent to DiD if there are more than two time periods in the panel design. This is not a concern in the current study with two time periods. Furthermore, exposure to political campaigns is a time-specific variable that could increase online political participation temporarily. But the variation in this exposure level between individuals is fairly small and therefore can be overlooked in this study.  Further, within-unit and within-time variations are not a concern here because as Imai & Kim (2021) show, in TWFE, the estimator is applied after within-time and within-unit variations are subtracted from the overall variation.

Simultaneity or reverse causality. Another empirical challenge in this study is the probability of reverse causality in establishing the relationship between online and offline political participation (H4). In their recent meta-analysis examining 89 coefficients across 30 studies, Boulianne et al. (2020) found that when the assumption of directionality is from offline activity to online participation, the coefficients are more likely to be positive and significant. However, most of the studies are based on cross-sectional data and cannot examine reverse causality. Even with longitudinal data, it is not always possible to show the direction of causality.

 To solve the issue of reverse causality, this study uses an instrumental variable approach in which the variation in online political participation that is caused by casual social media use is used to estimate the effect on offline political activity. Casual social media use is a strong predictor for online political participation, as is explained in the result section. Also, casual social media use is arguably unlikely to cause variation in offline political activity other than through online political participation, given the covariates.

## 2.4. Analytical models

Hypothesis 1. To test the first hypothesis, four different models were used: pooled OLS, OLS with the lagged outcome, panel fixed-effects, and panel random effects. Equation (1) shows the identification strategy to estimate the effect of casual social media use on online political activity (H1). In equation (1), i denotes an individual and y denotes the year. $ln_{iy}$ represents the online political participation in log scale and $SMU_{iy}$ shows the index of casual social media use for individual i in year y. $X_{iy}$ is a vector of time-varying observed control variables. The two-way fixed effects are captured in this model by $\sigma_i$ for unit-specific and $\theta_y$ for time-specific effects. Finally, $\varepsilon_{iy}$ is the individual-level error term. Control variables in this model are online social network size, whether the individual is the head of their household, their household size, income level, binary marriage status, binary employment status, and age.

$$ln_{iy} = \alpha + \beta_1 SMU_{iy} + \beta_2 X_{iy} + \sigma_i + \theta_y + \varepsilon_{iy} \quad (1)$$

Hypothesis 2. To test the first hypothesis, four different models were used: pooled OLS, OLS with lagged outcome, panel fixed-effects, and panel random-effects. The second model, shown in equation (2), estimates the effect of casual social media use on offline political activity. In this equation, $ln_{iy}$ represents offline political activity for individual i in year y in log scale. The other notations are the same as for equation (1). Drawing causal inference from this equation requires an assumption that there are no unobserved time-varying confounders. This assumption can be assessed by thinking about the underlying mechanisms for the effect of social media use on online political participation, which is related to time-saving, exposure to political information, and changes in a person's social network (Kim & Chen, 2016). All these factors are unit-specific (but time-invariant) and time-specific (but unit-invariant); therefore, this assumption can be released (Imai & Kim, 2021). Other assumptions that need to be true are that past treatments do not affect the current outcome, and past outcomes do not affect current treatment (Imai & Kim, 2019). The effect of past social media use on current political participation is unlikely to be direct. First, given the underlying mechanisms discussed earlier, the effect happens in short periods, and the cause (social media use) should be continuously available to have an effect. For example, if social media use saves an individual time or exposes them to political information, or encourages them through their social network, these changes could only affect the outcome (online political participation) in a short time. However, past social media use could affect the current online political participation through past online political participation. Since here we have only two time periods, in a tradeoff between causal dynamics and time-invariant unobservable, I choose to focus on time-invariant unobservables and rule out their effect by implementing a two-way fixed effects model.

$$ln_{iy} = \alpha + \beta_1 SMU_{iy} + \beta_2 X_{iy} + \sigma_i + \theta_y + \varepsilon_{iy} \quad (2)$$

Hypothesis 4. Four different models were used to test the first hypothesis: pooled OLS, OLS with the lagged outcome, panel fixed-effects, and panel random effects. Estimating the effect of online political participation on offline political activity by OLS or fixed effects methods are probably biased. The main concern here is the plausible simultaneity or reverse causality. This study uses casual social media use as an instrumental variable to estimate the causal effect of online political participation on offline political activity. To draw causal inferences using this approach, the instrumental variable must be independent of unobserved determinants of the outcome variable. This means that by conditioning the covariates, it is plausible to believe that casual social media use is unrelated to unobserved causes of offline political activity (Sovey et al., 2011). The second condition for the instrumental variable is exclusion criteria (Sovey et al., 2011). This assumption is examined by including the variable online political participation ($ln_{iy}$) in equation (2) and running the model. If the coefficient of social media use is close to zero, it would indicate that the exclusion criteria are satisfied.

$$ln_{iy} = \alpha + \beta_1 \hat{ln}_{iy} + \beta_2 X_{iy} + \sigma_i + \theta_y + \varepsilon_{iy} \quad (3)$$

The result of equation (1) shows the strength of the instrumental variable in the first stage. Equation (3) shows the second stage of fixed-effects regression. $\hat{ln}_{iy}$ represents the predicted value for online political participation for individual i in year y in the first-stage model. The monotonicity assumption is quite reasonable in this context because the treatment of online political participation is not available for those who do not engage in online political participation. And finally, the stable unit treatment value assumption holds here because the sample is drawn all across the US, and it is extremely unlikely to assume that a given individual in the sample is affected by treatments assigned to other individuals.

## 3. Results

### 3.1. Descriptive statistics

Table 2 shows the summary statistics of the studied cohort. The average age of this cohort in 2015 was about 21 and 45 percent of respondents are male. The index of casual social media use in the sample is smoothly distributed with an average value of 11.45 in 2013 and 11.38 in 2015. Both online and offline political participation indices have decreased from 2013 to 2015, which is consistent with the decrease in social media use.

*Table 2: Descriptive Statistics*

|  | Number of Observations | Mean | Standard Deviation | Min | Max |
|---|---|---|---|---|---|
| Year =2013 | | | | | |
| Age | 1033 | 20.967 | 3.912 | 15 | 27 |
| Highschool | 1033 | .699 | .459 | 0 | 1 |
| Male | 1033 | .449 | .498 | 0 | 1 |
| SMU | 1009 | 11.446 | 6.012 | 0 | 20 |
| log on pol | 978 | .784 | 1.026 | 0 | 3.219 |
| log off pol | 982 | .189 | .412 | 0 | 1.792 |
| zSMU | 1009 | .006 | .991 | -1.881 | 1.415 |
| net size | 1033 | .166 | .372 | 0 | 1 |
| on pol | 978 | 3.038 | 5.22 | 0 | 24 |
| off pol | 982 | .352 | .865 | 0 | 5 |
| Vote | 969 | .75 | .433 | 0 | 1 |
| Head of household | 1033 | .543 | .498 | 0 | 1 |
| H.SIZE | 1032 | 3.57 | 1.624 | 1 | 12 |
| Income | 1024 | 10.855 | 4.85 | 1 | 19 |
| Married | 1033 | .101 | .301 | 0 | 1 |
| Working | 1033 | .479 | .5 | 0 | 1 |
| Year= 2015 | | | | | |
| Age | 1033 | 20.967 | 3.912 | 15 | 27 |
| Highschool | 1033 | .699 | .459 | 0 | 1 |
| Male | 1033 | .449 | .498 | 0 | 1 |
| SMU | 1016 | 11.377 | 6.125 | 0 | 20 |
| ln on pol | 1013 | .688 | .977 | 0 | 3.219 |
| ln off pol | 1016 | .139 | .375 | 0 | 1.792 |
| zSMU | 1016 | -.006 | 1.009 | -1.881 | 1.415 |
| net size | 1033 | .198 | .399 | 0 | 1 |
| on pol | 1013 | 2.575 | 4.778 | 0 | 24 |
| off pol | 1016 | .271 | .822 | 0 | 5 |
| Vote | 971 | .577 | .494 | 0 | 1 |
| H.Head | 1033 | .558 | .497 | 0 | 1 |
| H.SIZE | 1028 | 3.535 | 1.597 | 1 | 12 |
| INCOME | 1005 | 11.021 | 4.887 | 1 | 19 |
| Married | 1033 | .136 | .343 | 0 | 1 |
| Working | 1033 | .646 | .479 | 0 | 1 |

SMU: Social Media Use, on pol: Online Political Participation, off pol: Offline Political Participation, zSMU: Standardized SMU, net size: Social Network Size, H.SIZE: Household Head, H.SIZE: Household Size.

### 3.2. Effect of Social Media Use on Online Political Participation

Table 3 shows the results of the four models used for testing the first hypothesis. All these models estimated a significant positive relationship (with different magnitudes) between social media use and online political participation. The pooled OLS model estimates 0.275 increase in online political participation with every one unit increase in social media use among the studied cohort. This estimation is biased due to confounding factors. In the second model that uses lagged outcome, a smaller coefficient of 0.236 was estimated. The result of the fixed-effects model that adjusts for unit and time-specific fixed effects shows the smallest effect size of 0.115. Finally, the result of the random-effects model with the same covariates as the fixed-effects model shows a larger coefficient than that of the fixed-effects model. Given the considerable difference between the effect sizes in the fixed-effects and random-effects models, and the fact that coefficients in pooled OLS and random-effects models are very close, it seems that the problem of confounder bias is a serious issue. Moreover, the Hausman and Mundlak's (1978) tests verify that the estimated effect size of the random-effects model is not consistent (Prob>chi2 = 0.0004 and Prob > F = 0.000 respectively), and the fixed effects model gives the most reliable estimate.

The result shows that for every one standard deviation increase in social media use for communication with friends and family, there is an 11.5 percent increase in the frequency of online political participation on average. This relationship is depicted in Figure 2 which shows the fixed-effects model's prediction of online political participation for different values of social media use. A strong prediction of online political participation by social media use also provides legitimacy for SMU to be used as an instrumental variable in equation (3). Moreover, the F-test of the excluded instrumental variable is greater than 10. Interestingly, findings do not show heterogeneity of effect regarding the network size binary variable. The effect size for individuals with a large online social network (more than 500 followers) is the same as the effect for individuals with small networks.

*Table 3: Results of the effect of SMU on Online Political Participation*

|  | (1) Pooled OLS | (2) OLS with Lagged Outcome | (3) Fixed Effects | (4) Random Effects |
|---|---|---|---|---|
| Standardized SMU | 0.275*** | 0.159*** | 0.115** | 0.268*** |
|  | (0.031) | (0.049) | (0.058) | (0.025) |
| Large Network | 0.180* | 0.156 | -0.023 | 0.153** |
|  | (0.094) | (0.109) | (0.099) | (0.060) |
| Large Network*SMU | 0.037 | 0.049 | -0.006 | 0.058 |
|  | (0.099) | (0.118) | (0.101) | (0.065) |
| Household Head | 0.034 | 0.118 | 0.078 | 0.035 |
|  | (0.068) | (0.076) | (0.122) | (0.051) |
| Household size | -0.017 | 0.010 | -0.019 | 0.002 |
|  | (0.018) | (0.023) | (0.048) | (0.014) |
| House income | -0.004 | 0.001 | 0.008 | -0.011** |
|  | (0.008) | (0.008) | (0.011) | (0.005) |
| Married | -0.000 | -0.010 | -0.105 | 0.011 |
|  | (0.109) | (0.108) | (0.165) | (0.082) |
| Working | -0.097 | -0.039 | 0.045 | -0.012 |
|  | (0.074) | (0.082) | (0.100) | (0.049) |
| Age | 0.007 | 0.007 |  | 0.014* |
|  | (0.011) | (0.010) |  | (0.008) |
| 2015 | -0.090* | -0.027 | -0.076 | -0.118*** |
|  | (0.048) | (0.077) | (0.047) | (0.036) |
| Lagged Online Participation |  | 0.419*** |  |  |
|  |  | (0.050) |  |  |
| _cons | 0.664** | 0.100 | 0.608** | 0.555*** |
|  | (0.266) | (0.296) | (0.237) | (0.200) |
| Obs. | 1930 | 933 | 1930 | 1930 |
| Pseudo R2 | .z | .z | .z | .z |

Standard errors are in parenthesis.
  * P<0.05, ** P<0.01,  *** P<0.001

*Figure 2: Predicted Margins of Online Political Participation Over SMU*

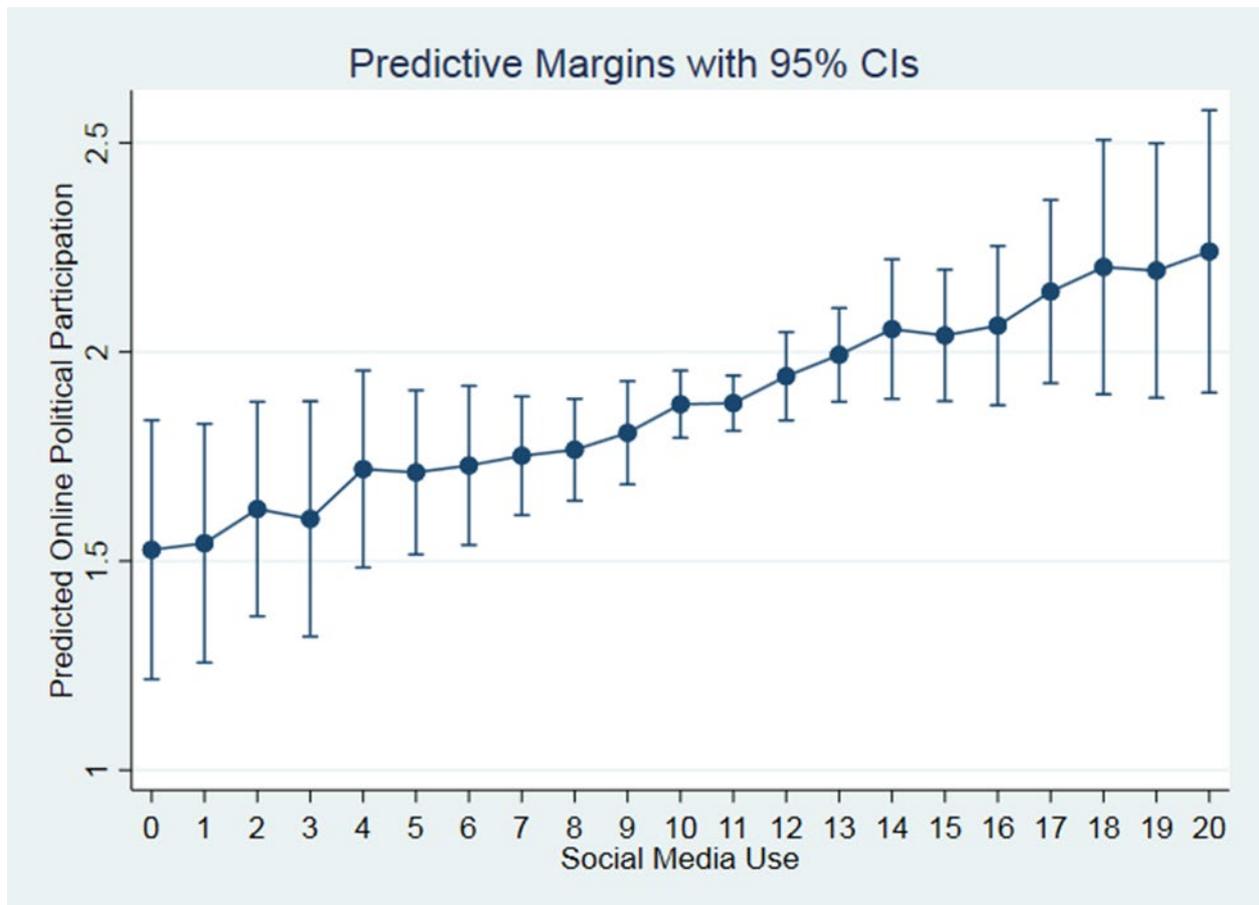

### 3.3. Effect of Social Media Use on Offline Political Activity

Table 4 shows the results of four regression models for estimating the effect of social media use on offline political activity. The pooled OLS model shows an effect size of around 4 percent for one standard deviation increase in casual social media use and it is statistically significant in the level of $p<0.01$. The estimated effect size resulting from the lagged outcome and random effect models are 3.6 and 3.2 percent, respectively, and both are statistically significant. However, the result of the fixed-effect model in the third column shows a negligible effect size of a 1.4 percent increase for one standard deviation increase in social media use, and more importantly, the coefficient is not statistically significant even at the level of $p<0.4$. Implementing both Hausman and Mundlak's (1978) tests shows that random effect estimates are inconsistent (Prob>chi2 = 0.0085 and Prob > F = 0.00, respectively). The inconsistency of results between fixed-effects on the one hand, and pooled OLS, lagged outcome, and random effect models, on the other hand, provides more evidence that the fixed-effects model reflects the true effect.

Table 4: Results for the effect of SMU on Offline Political Participation

|  | (1) Pooled OLS | (2) OLS Lagged Outcome | (3) Fixed Effects | (4) Random Effects |
|---|---|---|---|---|
| Standardized SMU | 0.040*** | 0.036** | 0.014 | 0.032*** |
|  | (0.011) | (0.017) | (0.018) | (0.010) |
| Large Network | 0.045 | 0.039 | 0.015 | 0.049* |
|  | (0.045) | (0.039) | (0.036) | (0.026) |
| Large Network*SMU | 0.075 | 0.043 | 0.026 | 0.047 |
|  | (0.047) | (0.051) | (0.035) | (0.031) |
| House Head | 0.015 | 0.030 | -0.006 | 0.020 |
|  | (0.027) | (0.029) | (0.034) | (0.021) |
| Household size | -0.011 | 0.001 | -0.007 | 0.001 |
|  | (0.007) | (0.008) | (0.023) | (0.006) |
| Household income | 0.001 | 0.004 | -0.003 | -0.005** |
|  | (0.003) | (0.003) | (0.005) | (0.002) |
| Married | -0.034 | -0.043 | -0.007 | -0.017 |
|  | (0.038) | (0.051) | (0.085) | (0.029) |
| Working | -0.017 | 0.001 | 0.081** | 0.022 |
|  | (0.028) | (0.031) | (0.040) | (0.019) |
| Age | 0.003 | 0.004 |  | 0.001 |
|  | (0.005) | (0.004) |  | (0.003) |
| 2015 | -0.056*** | -0.076** | -0.061*** | -0.053*** |
|  | (0.019) | (0.030) | (0.018) | (0.014) |
| Lagged Offline |  | 0.489*** |  |  |
|  |  | (0.060) |  |  |
| _cons | 0.127 | -0.038 | 0.194* | 0.187** |

|  | (0.115) | (0.118) | (0.100) | (0.083) |
|---|---|---|---|---|
| Obs. | 1936 | 941 | 1936 | 1936 |
| Pseudo R2 | .z | .z | .z | .z |

Standard errors are in parenthesis

*** p<0.01, ** p<0.05, * p<0.1

### 3.4. Effect of Social Media Use on Voting

Table 5 shows the effect of social media use on the probability of voting in the youth. The estimated effects in the pooled logit and the random effect logit are significant at the level of p<0.01. However, the effect size from the fixed-effect logit model is only significant at p<0.1.

*Table 5: Results of the effect of SMU on Offline Voting*

|  | (1) | (2) | (3) |
|---|---|---|---|
|  | Pooled Logit | Fixed-effects _Logit | Random-effects _Logit |
| Standardized SMU | 1.211*** | 1.360* | 1.439*** |
|  | (0.067) | (0.226) | (0.150) |
| Large Network | 1.058 | 0.890 | 1.179 |
|  | (0.153) | (0.395) | (0.308) |
| Large Network*SMU | 0.941 | 0.453* | 0.781 |
|  | (0.145) | (0.213) | (0.212) |
| 2015 | 0.462*** | 0.231*** | 0.250*** |
|  | (0.048) | (0.042) | (0.040) |
| House Head | 1.286** | 1.095 | 1.439* |
|  | (0.134) | (0.451) | (0.288) |
| House size | 0.979 | 1.087 | 0.964 |
|  | (0.032) | (0.156) | (0.060) |
| House income | 1.045*** | 0.906** | 1.062*** |

|  | (0.011) | (0.044) | (0.022) |
| --- | --- | --- | --- |
| Married | 0.909 | 0.609 | 0.832 |
|  | (0.147) | (0.425) | (0.261) |
| Working | 0.819* | 0.756 | 0.681** |
|  | (0.086) | (0.231) | (0.130) |
| _cons | 1.958*** |  | 4.458*** |
|  | (0.379) |  | (1.722) |
| /lnsig2u |  |  | 5.852*** |
|  |  |  | (1.055) |
| Obs. | 1875 | 492 | 1875 |
| Pseudo R2 | 0.044 | 0.325 | .z |

Standard errors are in parenthesis
*** p<0.01, ** p<0.05, * p<0.1

### 3.5. Effect of Online Political Activity on Offline Political Activity

Table 6 shows the results of four specification models for the estimation of the effect of online political participation on offline political activity. Surprisingly, all four coefficients are statistically significant at the level of p<0.01, and all are positive and consistently large. The effect sizes resulting from pooled OLS and pooled IV models (columns 1 and 3 in table 6) are around 0.16. This means that a one percent increase in online political participation translates to a 0.16 percent increase in offline political activity in the youth. The results from fixed effects and IV fixed-effect models (columns 2 and 4 in table 6) show a coefficient value of around 0.12 which implies that a one percent increase in online political participation leads to a 0.12 percent increase in offline political activity.

*Table 6: Results for the effect of Online Political Participation on Offline Political Activity*

|  | (1) | (2) | (3) | (4) |
| --- | --- | --- | --- | --- |
|  | Pooled OLS | Fixed Effects | Pooled IV | Fixed Effects IV |
| Log Online Participation | 0.160*** | 0.117*** | 0.154*** | 0.125*** |
|  | (0.020) | (0.027) | (0.039) | (0.024) |
| SMU | -0.001 | 0.005 |  |  |
| | (0.010) | (0.016) | | |
| Large Network | -0.059** | 0.008 | -0.063** | -0.008 |
| | (0.025) | (0.034) | (0.031) | (0.040) |
| Large Network*SMU | 0.111** | 0.015 | 0.118** | |
| | (0.053) | (0.041) | (0.054) | |
| House Head | 0.009 | -0.006 | | |
| | (0.023) | (0.033) | | |
| House size | -0.010* | -0.006 | -0.011** | 0.013 |
| | (0.006) | (0.020) | (0.005) | (0.011) |
| House income | 0.001 | -0.003 | 0.000 | -0.005 |
| | (0.003) | (0.004) | (0.002) | (0.004) |
| Married | -0.040 | -0.005 | -0.033 | 0.011 |
| | (0.034) | (0.083) | (0.028) | (0.051) |
| Working | 0.005 | 0.082** | 0.008 | 0.059** |
| | (0.025) | (0.035) | (0.022) | (0.025) |
| Age | 0.002 | -0.046* | | |
| | (0.004) | (0.024) | | |
| 2015 | -0.038** | 0.034 | -0.035* | -0.043*** |
| | (0.018) | (0.046) | (0.020) | (0.014) |
| _cons | 0.036 | 1.074** | 0.086 | 0.081 |
| | (0.096) | (0.517) | (0.054) | (0.061) |
| Obs. | 1897 | 1897 | 1897 | 1897 |
| Pseudo R2 | .z | .z | .z | .z |

Standard errors are in parenthesis
*** $p<0.01$, ** $p<0.05$, * $p<0.1$

## 4. Discussion

With the rapid increase in the use of social media in the younger population, it is imperative to understand how it influences their political participation. Previous studies have investigated the correlations between social media use and political participation, but evidence of a causal relationship is rare. This study aimed to investigate the causal link between the casual use of social media and online and offline political participation in the youth.

This study did not find evidence of a direct effect of casual social media use on offline political activities. This means that contrary to the findings of the majority of previous studies (Boulianne & Theocharis, 2020; Gil de Zúñiga, 2012) and consistent with the result from (Kahne & Bowyer, 2018), social media use does not have a significant effect on the offline political activities in the youth. However, the current study's findings suggest an indirect effect of casual social media use on offline political activities through online political participation. These findings challenge the ongoing claims about slacktivism in youth, which suggest that activity on social media does not necessarily substitute offline activities. Given the prevalence of social media use among youth and the increasing frequency of their usage, these findings predict a more politically engaged generation through the coming years. The presence of the right conditions might foster democratic processes around the world.

These results provide strong evidence against the idea of slacktivism (Twenge, 2017) which claims that online engagement with politics in the youth does not lead to offline political activities but rather substitutes them. The lack of difference between models with and without instrumental variables has implications in the debates about the directionality of effects (Boulianne & Theocharis, 2020). These results provide evidence that the direction of causality is from online political participation to offline political activities and not the other way around.

## 5. Conclusion

This study has several limitations, which suggest areas for future research. First, the survey data was collected in 2015. With the rapid development of social media use and the growth of the new youth generation, repeating the survey in the future will be valuable in improving our understanding of the effects of social media use in the younger population. This study provides an example of how such data can be used in the future. Second, the availability of new data on this cohort can improve the power of analyses used in this study for causal inference. With data for more time periods, there would have been more within-variation in the variables of interest, which increases the reliability of the results. Third, the Likert scale method used for measuring the variables of interest has raised some concerns. Debates are still going on around whether a Likert score is a proper measure of social media use and online and offline political engagement. More recent methods such as web metrics, click metrics, Click Through Rate (CTR), web scraping, text analysis, and machine learning methods could be implemented to gain more reliable and analysis-friendly measures of social media use and online political participation (Barati & Ansari, 2022). These methods extract information from the daily activities of online users, which might provide an unobtrusive insight into the youth's online political behavior. However, using these methods raises significant legal, ethical, and privacy concerns, which require careful attention from researchers.

The use of online participation as an instrumental variable has some limitations. First, there could be potential causal pathways between online participation and offline participation (such as individual personality traits), which might violate the exclusion criteria of an instrumental variable. However, this is unlikely because mere casual use of social media (for family and friend communications) cannot cause any sort of involvement in offline political participation without first activating at least some sort of online political participation. More research is needed to examine these pathways. Second, it is possible that some people participate in online political activities but not offline (e.g., clicktivism), which might violate the monotonicity assumption of instrumental variables. However, because clicktivism is a recent phenomenon, it is unlikely that this is a significant problem in the 2014 data used in the study. Finally, there were some limitations in the data we used, such as a need for contextual variables such as the reason for participation. However, we maintain that this issue would not change the results significantly.

## References

Allcott, H., Braghieri, L., Eichmeyer, S., & Gentzkow, M. (2020). The Welfare Effects of Social Media. American Economic Review, 110(3), 629–676. https://doi.org/10.1257/aer.20190658

Anderson, M., & Jiang, J. (2018, May 31). Teens, Social Media and Technology 2018. Pew Research Center: Internet, Science & Tech. https://www.pewresearch.org/internet/2018/05/31/teens-social-media-technology-2018/

Auxier, B., & Anderson, M. (2021, April 7). Social Media Use in 2021. Pew Research Center: Internet, Science & Tech. https://www.pewresearch.org/internet/2021/04/07/social-media-use-in-2021/

Bailey, M. A. (2016). Real Stats: Using Econometrics for Political Science and Public Policy. Oxford University Press.

Barati, M., & Ansari, B. (2022). Effects of algorithmic control on power asymmetry and inequality within organizations. Journal of Management Control, 0123456789. https://doi.org/10.1007/s00187-022-00347-6

Boulianne, S. (2015). Social media use and participation: A meta-analysis of current research. Information Communication and Society, 18(5), 524–538. https://doi.org/10.1080/1369118X.2015.1008542

Boulianne, S., & Theocharis, Y. (2020). Young People, Digital Media, and Engagement: A Meta-Analysis of Research. Social Science Computer Review, 38(2), 111–127. https://doi.org/10.1177/0894439318814190

Carr, C. T., & Hayes, R. A. (2015). Social Media: Defining, Developing, and Divining. Atlantic Journal of Communication, 23(1), 46–65. https://doi.org/10.1080/15456870.2015.972282

Cohen, C., & Kahne, J. (2018). Youth Participatory Politics Survey Project, United States. https://doi.org/10.3886/ICPSR37188.v1

Gil de Zúñiga, H. (2012). Social Media Use for News and Individuals' Social Capital, Civic Engagement and Political Participation. Journal of Computer-Mediated Communication, 17(3), 319–336. https://doi.org/10.1111/j.1083-6101.2012.01574.x

Imai, K., & Kim, I. S. (2019). When Should We Use Unit Fixed Effects Regression Models for Causal Inference with Longitudinal Data? American Journal of Political Science, 63(2), 467–490. https://doi.org/10.1111/ajps.12417

Imai, K., & Kim, I. S. (2021). On the Use of Two-Way Fixed Effects Regression Models for Causal Inference with Panel Data. Political Analysis, 29(3), 405–415. https://doi.org/10.1017/pan.2020.33

Jenkins, H., Ford, S., & Green, J. (2018). Spreadable media: Creating value and meaning in a networked culture.

Kahne, J., & Bowyer, B. (2018). The Political Significance of Social Media Activity and Social Networks. Political Communication, 35(3), 470–493. https://doi.org/10.1080/10584609.2018.1426662

Kim, Y., & Chen, H. T. (2016). Social media and online political participation: The mediating role of exposure to cross-cutting and like-minded perspectives. Telematics and Informatics, 33(2), 320–330. https://doi.org/10.1016/j.tele.2015.08.008

Knoll, J., Matthes, J., & Heiss, R. (2020). The social media political participation model: A goal systems theory perspective. Convergence: The International Journal of Research into New Media Technologies, 26(1), 135–156. https://doi.org/10.1177/1354856517750366

Kross, E., Verduyn, P., Demiralp, E., Park, J., Lee, D. S., Lin, N., Shablack, H., Jonides, J., & Ybarra, O. (2013). Facebook Use Predicts Declines in Subjective Well-Being in Young Adults. PLoS ONE, 8(8), 1–6. https://doi.org/10.1371/journal.pone.0069841

Lorenz-Spreen, P., Oswald, L., Lewandowsky, S., & Hertwig, R. (2022). A systematic review of worldwide causal and correlational evidence on digital media and democracy. Nature Human Behaviour, 1–28. https://doi.org/10.1038/s41562-022-01460-1

Obar, J. A., & Wildman, S. S. (2015). Social media definition and the governance challenge-an introduction to the special issue. Obar, JA and Wildman, S.(2015). Social Media Definition and the Governance Challenge: An Introduction to the Special Issue. Telecommunications Policy, 39(9), 745–750.

Pew Research Center. (2021, April 7). Social Media Fact Sheet. Pew Research Center: Internet, Science & Tech. https://www.pewresearch.org/internet/fact-sheet/social-media/

Sovey, A. J., Green, D. P., Sovey, A. J., & Green, D. P. (2011). Instrumental Variables Estimation in Political Science: A Readers' Guide. American Journal of Political Science, 55(1), 188–200. https://doi.org/10.1111/j.l540-5907.2010.00477.x

Twenge, J. M. (2017). IGen: Why today's super-connected kids are growing up less rebellious, more tolerant, less happy—And completely unprepared for adulthood—And what that means for the rest of us. Simon and Schuster.

Vitak, J., Zube, P., Smock, A., Carr, C. T., Ellison, N., & Lampe, C. (2011). It's complicated: Facebook users' political participation in the 2008 election. Cyberpsychology, Behavior and Social Networking, 14(3), 107–114. https://doi.org/10.1089/cyber.2009.0226

Xenos, M., Vromen, A., & Loader, B. D. (2014). The great equalizer? Patterns of social media use and youth political engagement in three advanced democracies. Information Communication and Society, 17(2), 151–167. https://doi.org/10.1080/1369118X.2013.871318

## About the Author

*Mehdi Barati*

Mehdi Barati is a Ph.D. candidate in information science at the State University of New York at Albany, New York, United States. He conducts research at the intersection of data analytics and work organization, focusing on the use of artificial intelligence in organizations for workforce management.